\documentclass[11pt,a4paper]{article}
\pdfoutput=1
\usepackage{jheppub}
\usepackage{tikz}
\usetikzlibrary{intersections, calc, arrows.meta, bending}
\usepackage{subcaption}
\usepackage{simpler-wick}

\DeclareMathOperator{\Tr}{Tr}
\DeclareMathOperator{\tr}{tr}

\newcommand{\ri}{\mathrm{i}}
\renewcommand{\th}{\theta}

\newcommand{\vep}{\varepsilon}

\newcommand{\hf}{\frac{1}{2}}

\newcommand{\til}[1]{\widetilde{#1}}

\newcommand{\del}{\partial}

\newcommand{\bra}{\langle}
\newcommand{\ket}{\rangle}
\newcommand{\la}{\lambda}

\newcommand{\bt}{\beta}
\newcommand{\ga}{\gamma}

\newcommand{\al}{\alpha}
\newcommand{\om}{\omega}

\newcommand{\rt}[1]{\sqrt{#1}}
\newcommand{\cO}{\mathcal{O}}
\newcommand{\cZ}{\mathcal{Z}}

\newcommand{\cN}{\mathcal{N}}

\newcommand{\cJ}{\mathcal{J}}

\newcommand{\fb}{\mathfrak{b}}

\makeatletter
\gdef\@fpheader{}
\makeatother

\begin{document}
\title{de Sitter JT gravity from double-scaled SYK}

\author{Kazumi Okuyama}

\affiliation{Department of Physics, 
Shinshu University, 3-1-1 Asahi, Matsumoto 390-8621, Japan}

\emailAdd{kazumi@azusa.shinshu-u.ac.jp}

\abstract{
It is known that the double-scaled SYK model
(DSSYK) reduces to JT gravity with a negative 
cosmological constant by zooming in on the lower edge
$E=-E_0$
of the spectrum.
We find that the de Sitter JT gravity
(i.e. JT gravity with a positive 
cosmological constant) is reproduced from
DSSYK by taking a scaling limit around
the upper edge $E=E_0$ of the spectrum. 
We also argue that the appearance of de Sitter
JT gravity is consistent with the behavior of the
classical
solution of the sine dilaton gravity.
}

\maketitle

\section{Introduction}
Construction of de Sitter vacua in string theory
is a notoriously difficult problem (see e.g.
\cite{Obied:2018sgi,Berglund:2022qsb,Schachner:2025vol} and references
therein).
It might be the case that quantum gravity
in de Sitter space is intrinsically unstable \cite{Rajaraman:2016nvv,Polyakov:2012uc} and defined only as a resonance of flat space
S-matrix \cite{Maltz:2016iaw}. 
Also, the holography in de Sitter space is poorly understood
\cite{Strominger:2001pn} compared to its AdS cousin 
\cite{Witten:1998qj}.
It is highly desirable to find a 
tractable example of de Sitter 
quantum gravity.

The two-dimensional Jackiw-Teitelboim (JT)
gravity \cite{Jackiw:1984je,Teitelboim:1983ux} 
with a positive cosmological constant is a useful
toy model of de Sitter quantum gravity  \cite{Maldacena:2019cbz,Cotler:2024xzz,Cotler:2019nbi,Cotler:2019dcj,Cotler:2023eza,Moitra:2022glw,Nanda:2023wne,Dey:2025osp,Held:2024rmg,Alonso-Monsalve:2024oii,Collier:2025pbm}.
Another useful example is the so-called double-scaled SYK
model (DSSYK) \cite{Berkooz:2018jqr}.
It is argued that DSSYK is holographically dual to
de Sitter space in a certain high temperature limit
\cite{Susskind:2021esx,Susskind:2022dfz,Susskind:2022bia,Susskind:2023hnj,Rahman:2023pgt,Rahman:2024iiu,Sekino:2025bsc,Narovlansky:2023lfz,Verlinde:2024znh,Verlinde:2024zrh,Tietto:2025oxn}.
In \cite{Blommaert:2024whf}, 
another realization of de Sitter space
was proposed based on a duality between DSSYK and the sine
dilaton gravity \cite{Blommaert:2024ymv}.

In this paper, we will show that
the matrix model of de Sitter JT gravity
studied in \cite{Maldacena:2019cbz,Cotler:2024xzz}
is obtained from the ETH matrix model of DSSYK \cite{Jafferis:2022wez}
by taking a certain triple scaling limit around the upper end
$E=E_0$ of the eigenvalue spectrum.
At the disk level, the eigenvalue density $\rho_0(E)$
of DSSYK has a finite range of support $|E|\leq E_0$
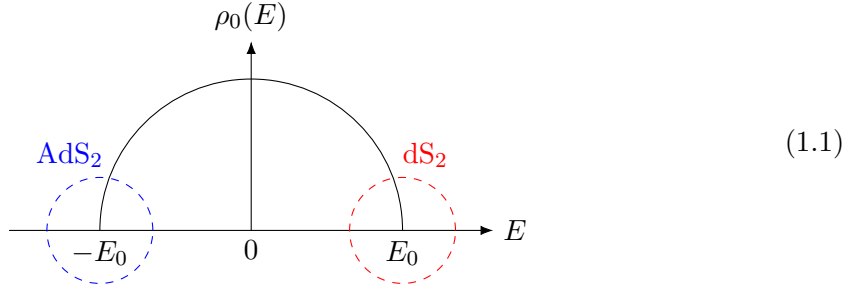
\begin{equation}
\begin{aligned}
 \begin{tikzpicture}[scale=1]
\draw (2,0) arc [start angle=0,end angle=180,radius=2]; 
\draw[-{Latex}] (-3.2,0)--(3.2,0); 
\draw[-{Latex}] (0,0)--(0,2.5); 
\draw (-2,0) node [below]{$-E_0$};
\draw (2,0) node [below]{$E_0$};
\draw (0,0) node [below]{$0$};
\draw (3.2,0) node [right]{$E$};
\draw (0,2.5) node [above]{$\rho_0(E)$};
\draw[dashed,blue] (-2,0) circle [radius=0.7];
\draw[dashed,red] (2,0) circle [radius=0.7];
\draw (-2.4,0.7) node [above] {\textcolor{blue}{AdS${}_2$}};
\draw (2.3,0.7) node [above] {\textcolor{red}{dS${}_2$}};
\end{tikzpicture}
\end{aligned} 
\label{eq:rho-fig}
\end{equation}
As shown in \cite{Berkooz:2018jqr,Lin:2022rbf},
DSSYK reduces to JT gravity with a negative 
cosmological constant by zooming in on the lower edge $E=-E_0$
of the spectrum.
In \cite{Blommaert:2024whf}, it is suggested that
the upper end $E=E_0$ of the spectrum corresponds to
the de Sitter JT gravity.
We will directly verify this claim by taking a scaling
limit of the ETH matrix model near $E=E_0$
and find a complete agreement with the matrix model
of de Sitter JT gravity
studied in \cite{Maldacena:2019cbz,Cotler:2024xzz}.
It turns out that the upper end $E=E_0$
of the spectrum 
is an unstable saddle point, which is consistent with
the picture that the quantum gravity on de Sitter space is
intrinsically unstable.

This paper is organized as follows.
In section \ref{sec:review},
we review the known results of DSSYK and the ETH matrix 
model.
In section \ref{sec:AdS-JT},
we review the fact that the scaling limit of ETH matrix model
around the lower edge $E=-E_0$
reproduces the matrix model of JT gravity in \cite{Saad:2019lba}.
In section \ref{sec:dS-JT},
we take a scaling limit of the ETH matrix model
around the upper edge $E=E_0$ and find
that the matrix model of de Sitter JT gravity \cite{Maldacena:2019cbz,Cotler:2024xzz} is
correctly reproduced.
In section \ref{sec:sine-dilaton},
we consider classical solutions of the sine dilaton gravity
and find that the metric of the 
solution corresponding to the upper edge
$E=E_0$ is minus the metric of $AdS_2$, which can be interpreted as 
a two-dimensional de Sitter space \cite{Maldacena:2019cbz}.
Finally,
we conclude in section \ref{sec:conclusion}
with some discussion on the future problems.

\section{Review of DSSYK and ETH matrix model}\label{sec:review}
In this section, we will briefly review the double-scaled SYK model (DSSYK)
and the ETH matrix model.
The Sachdev-Ye-Kitaev (SYK) model is a system of $N$ Majorana fermions $\psi_i~(i=1,\cdots,N)$ 
with all-to-all $p$-body interaction \cite{Sachdev1993,Kitaev1,Kitaev2}
\begin{equation}
\begin{aligned}
 H_\text{SYK}=\ri^{p/2}\sum_{i_i<\cdots <i_p}J_{i_1\dots i_p}\psi_{i_1}\cdots\psi_{i_p},
\end{aligned} 
\end{equation}
where $J_{i_1\dots i_p}$ is a Gaussian random coupling with zero mean and the variance
\begin{equation}
\begin{aligned}
 \bra J_{i_1\dots i_p}^2\ket=\cJ^2\binom{N}{p}^{-1}.
\end{aligned} 
\end{equation}
DSSYK is defined by the scaling limit
\begin{equation}
\begin{aligned}
 N,p\to \infty\quad\text{with}\quad \la=\frac{2p^2}{N}:~\text{fixed}.
\end{aligned} 
\end{equation}
In this limit, the computation of the moment $\bra\tr (H_\text{SYK})^k\ket$ 
boils down
to the counting problem of the intersection numbers of chord diagrams.
This counting problem can be exactly solved in terms of the $q$-deformed oscillators
with $q=e^{-\la}$.
In this way, we arrive at the exact result of the 
disk partition function of DSSYK \cite{Berkooz:2018jqr}
\begin{equation}
\begin{aligned}
 \cZ(\bt)=\int_0^\pi\frac{d\th}{2\pi}\mu(\th)
e^{-\bt E(\th)},
\end{aligned} 
\label{eq:disk}
\end{equation}
where $E(\th)$ and $\mu(\th)$ are given by
\begin{equation}
\begin{aligned}
 E(\th)=-E_0\cos\th,
\quad \mu(\th)=(q,e^{\pm2\ri\th};q)_\infty,
\end{aligned} 
\label{eq:Eth-muth}
\end{equation}
with $E_0=2\cJ/\rt{1-q}$.
Note that the disk partition function \eqref{eq:disk} is an even function
of $\bt$
\begin{equation}
\begin{aligned}
 \cZ(-\bt)=\cZ(\bt),
\end{aligned} 
\end{equation} 
which follows from the symmetry 
\begin{equation}
\begin{aligned}
 E(\pi-\th)=-E(\th),\quad \mu(\pi-\th)=\mu(\th).
\end{aligned} 
\label{eq:Emu-sym}
\end{equation}
An important property of DSSYK is that 
the energy spectrum $E(\th)$ is bounded from below and above
\begin{equation}
\begin{aligned}
 -E_0\leq E(\th)\leq E_0.
\end{aligned} 
\end{equation}
The lower end $E=-E_0$ and the upper end $E=E_0$ of the spectrum correspond
to $\th=0$ and $\th=\pi$, respectively.
The end-points $E=\pm E_0$ are characterized as
the saddle points of the Boltzmann factor
$e^{-\bt E(\th)}$
\begin{equation}
\begin{aligned}
 \frac{\del}{\del\th}E(\th)=0\quad\Rightarrow\quad \th=0,\pi.
\end{aligned} 
\end{equation}
It turns out that $\th=0$ is a stable saddle point while $\th=\pi$
is an unstable saddle point, as we will see below.

\subsection{ETH matrix model}
As discussed in \cite{Jafferis:2022wez},
one can introduce the $L\times L$ hermitian matrix model, 
called ETH matrix model, which reproduces the 
density of states $\mu(\th)$ of DSSYK in the large $L$ limit, 
\begin{equation}
\begin{aligned}
 \cZ=\int_{L\times L}dH e^{-L\Tr V(H)},
\end{aligned} 
\label{eq:ETH}
\end{equation}
where $L$ is the dimension of the Hilbert space of $N$ Majorana fermions
\begin{equation}
\begin{aligned}
 L=2^{N/2}.
\end{aligned} 
\end{equation}
By matching the genus-zero eigenvalue density $\rho_0(E)$ of ETH matrix model
with the density of states $\mu(\th)$ of DSSYK in \eqref{eq:Eth-muth}
\begin{equation}
\begin{aligned}
 \rho_0(E(\th))dE(\th)=\mu(\th)\frac{d\th}{2\pi},
\end{aligned} 
\end{equation}
the potential $V(H)$ in \eqref{eq:ETH} is determined as 
\begin{equation}
\begin{aligned}
V(H)&=\sum_{n=1}^\infty\frac{(-1)^{n-1}}{n}q^{\hf n^2}(q^{\hf n}+q^{-\hf n})
T_{2n}(H/E_0),
\end{aligned} 
\label{eq:VH}
\end{equation}
where $T_n(x)$ denotes the Chebyshev polynomial of the first kind.
Then the disk partition function $\cZ(\bt)$ of
DSSYK in \eqref{eq:disk} is reproduced
from the ETH matrix model \eqref{eq:ETH} in the large $L$ limit
\begin{equation}
\begin{aligned}
 \cZ(\bt)=\lim_{L\to\infty}\frac{1}{L}\bra\Tr e^{-\bt H}\ket,
\end{aligned} 
\end{equation}
where the expectation value is defined by
\begin{equation}
\begin{aligned}
 \bra f(H)\ket=\frac{1}{\cZ}\int dH e^{-L\Tr V(H)}f(H).
\end{aligned} 
\end{equation}
We can also consider the connected correlator
of $\Tr e^{-\bt H}$ in the ETH matrix model, and its
genus expansion
in the large $L$ limit is written as
\begin{equation}
\begin{aligned}
 \left\bra
\prod_{i=1}^n \Tr e^{-\bt_i H}\right\ket_{\text{conn}}
=\sum_{g=0}^\infty L^{2-2g-n}\cZ_{g,n}(\bt_1,\cdots,\bt_n).
\end{aligned} 
\label{eq:cZ-gn}
\end{equation}
In this notation, the disk partition function
$\cZ(\bt)$ in \eqref{eq:disk} is the $(g,n)=(0,1)$ contribution in \eqref{eq:cZ-gn}
\begin{equation}
\begin{aligned}
 \cZ(\bt)=\cZ_{0,1}(\bt).
\end{aligned} 
\end{equation}

We can systematically compute $\cZ_{g,n}$ in \eqref{eq:cZ-gn}
by the topological recursion 
with the spectral curve $(x(z),y(z))$ \cite{Okuyama:2023kdo}
\begin{equation}
\begin{aligned}
 x(z)&=\frac{E_0}{2}(z+z^{-1}),\\
y(z)&=\frac{1}{E_0}(z^{-1}-z)\prod_{n=1}^\infty
(1-q^n)(1-z^2q^n)(1-z^{-2}q^n).
\end{aligned} 
\end{equation}
Note that $x(z)$ is related to the energy spectrum $E(\th)$ in \eqref{eq:Eth-muth}
by
\begin{equation}
\begin{aligned}
 E(\th)=-x(e^{\ri\th}).
\end{aligned} 
\label{eq:E-x}
\end{equation}
Introducing $\om_{g,n}$ by
\begin{equation}
\begin{aligned}
 \left\bra\prod_{i=1}^n\Tr\frac{dx(z_i)}{x(z_i)+H}\right\ket_{\text{conn}}
=\sum_{g=0}^\infty L^{2-2g-n}\om_{g,n}(z_1,\cdots,z_n),
\end{aligned} 
\end{equation}
$\om_{g,n}$ can be computed by
the Eynard-Orantin's topological recursion \cite{Eynard:2007kz,Eynard:2008we}
\begin{equation}
\begin{aligned}
 \om_{g,n+1}(z_0,J)=\sum_{\al=\pm1}\underset{z=\al}{\text{Res}}K(z_0,z)
\Bigl[\om_{g-1,n+2}(z,z^{-1},J)+\sum_{h=0}^g\sum'_{I\subset J}
\om_{h,1+|I|}(z,I)\om_{g-h,1+n-|I|}(z^{-1},J\backslash I)\Bigr]
\end{aligned} 
\label{eq:top-rec}
\end{equation}
together with the initial conditions
\begin{equation}
\begin{aligned}
\om_{0,1}(z)=-y(z)dx(z),\quad
\om_{0,2}(z_1,z_2)=\frac{dz_1dz_2}{(z_1-z_2)^2}.
\end{aligned} 
\end{equation}
The recursion kernel $K(z_0,z)$ is given by
\begin{equation}
\begin{aligned}
 K(z_0,z)&=\frac{\int_{z'=z^{-1}}^z \om_{0,2}(z_0,z')}{4y(z)dx(z)}.
\end{aligned} 
\end{equation}
In \eqref{eq:top-rec}, the prime in the summation means that
$(h,I)=(0,\emptyset),(g,J)$ are excluded.
Note that in \eqref{eq:top-rec} we take the residue at $z=\pm1$, which 
correspond to the branch points of $x(z)$
\begin{equation}
\begin{aligned}
 \frac{dx(z)}{dz}=0\quad \Rightarrow \quad z=\pm1.
\end{aligned} 
\end{equation}
From the relation \eqref{eq:E-x}, 
$z=\pm1$ correspond to $E=\mp E_0$.

As discussed in \cite{norbury2013polynomials},
$\om_{g,n}$ can be expanded as
\begin{equation}
\begin{aligned}
 \om_{g,n}(z_1,\cdots,z_n)=\sum_{\fb_1,\cdots,\fb_n\in\mathbb{Z}_{+}}
N_{g,n}(\fb_1,\cdots,\fb_n)\prod_{i=1}^n \fb_iz_i^{\fb_i-1}dz_i,
\end{aligned} 
\label{eq:om-N}
\end{equation}
where $N_{g,n}(\fb_1,\cdots,\fb_n)$ is  interpreted as a discrete analogue
of the volume of the moduli space of Riemann surfaces
of genus $g$ with $n$ boundaries.
Finally, $\cZ_{g,n}$ in \eqref{eq:cZ-gn} is obtained from $\om_{g,n}$ as
\begin{equation}
\begin{aligned}
 \cZ_{g,n}(\bt_1,\cdots,\bt_n)&=
\left(\prod_{i=1}^n\oint_{z_i=0}\frac{e^{\bt_i x(z_i)}}{2\pi\ri}\right)
\om_{g,n}(z_1,\cdots,z_n)\\
&=\sum_{\fb_1,\cdots,\fb_n\in\mathbb{Z}_{+}}
N_{g,n}(\fb_1,\cdots,\fb_n)\prod_{i=1}^n \fb_i
\cZ_{\text{trumpet}}(\bt_i,\fb_i),
\end{aligned}
\label{eq:Z-zint}
\end{equation}
where $\cZ_{\text{trumpet}}(\bt,\fb)$ 
is the trumpet partition function of DSSYK \cite{Jafferis:2022wez,Okuyama:2023byh}
\begin{equation}
\begin{aligned}
 \cZ_{\text{trumpet}}(\bt,\fb)=
\oint_{z=0}\frac{dz}{2\pi\ri z}z^{\fb} e^{\bt x(z)}
=I_{\fb}(\bt E_0).
\end{aligned} 
\label{eq:trumpet-DSSYK}
\end{equation}
Here $I_\nu(x)$ denotes the modified Bessel function of the first kind.

\section{Scaling limit around $E=-E_0$}\label{sec:AdS-JT}

\subsection{Matrix model of JT gravity}
As shown in \cite{Saad:2019lba}, 
JT gravity with a negative cosmological constant
is equivalent to a double-scaled matrix model.
At the disk level, the eigenvalue density $\rho_0(E)$
of JT gravity matrix model
is given by 
\begin{equation}
\begin{aligned}
 \rho_0(E)=\frac{\ga}{2\pi^2}\sinh(2\pi\rt{2\ga E}),
\end{aligned} 
\end{equation}
and the disk partition function of JT gravity becomes
\begin{equation}
\begin{aligned}
 Z_{\text{JT}}(\bt)=\int_0^\infty dE\rho_0(E)e^{-\bt E}
=\frac{1}{\rt{2\pi(\bt/\ga)^3}}e^{\frac{2\pi^2\ga}{\bt}},
\end{aligned} 
\label{eq:Z-JT}
\end{equation}
where $\ga$ is the boundary value of the dilaton.
In the following discussion, 
it is convenient to change the integration variable as
\begin{equation}
\begin{aligned}
E=\frac{k^2}{2\ga}. 
\end{aligned} 
\end{equation}
Then $Z_{\text{JT}}(\bt)$ in \eqref{eq:Z-JT}
is written as
\begin{equation}
\begin{aligned}
 Z_{\text{JT}}(\bt)=\int_{-\infty}^\infty\frac{dk}{2\pi}
\rho_s(k)e^{-\frac{\bt}{2\ga}k^2},
\end{aligned} 
\label{eq:Z-JT2}
\end{equation}
where $\rho_s(k)$ is the Schwarzian density of states
\cite{Stanford:2017thb}
\begin{equation}
\begin{aligned}
 \rho_s(k)=\frac{k\sinh(2\pi k)}{2\pi}.
\end{aligned}
\label{eq:rho-s} 
\end{equation}

More generally, the connected correlator of $Z(\bt)$ admits a genus expansion
\begin{equation}
\begin{aligned}
 \bra Z(\bt_1)\cdots Z(\bt_n)\ket_{\text{conn}}=\sum_{g=0}^\infty
e^{-(2g-2+n)S_0}Z_{g,n}(\bt_1,\cdots,\bt_n).
\end{aligned} 
\end{equation}
Note that the disk partition function \eqref{eq:Z-JT}
corresponds to $(g,n)=(0,1)$
\begin{equation}
\begin{aligned}
 Z_{\text{JT}}(\bt)=Z_{0,1}(\bt)
=\int_{C}\frac{dz}{2\pi\ri}W_{0,1}(z)e^{\frac{\bt}{2\ga}z^2},
\end{aligned} 
\label{eq:Z01}
\end{equation}
where the contour $C=\ri\mathbb{R}$ is along the imaginary $z$-axis and
$W_{0,1}(z)$ is given by
\begin{equation}
\begin{aligned}
 W_{0,1}(z)=-2z y(z),\quad y(z)=\frac{\sin(2\pi z)}{4\pi}.
\end{aligned} 
\label{eq:y-sin}
\end{equation}
The above expression \eqref{eq:Z01} can be generalized to
arbitrary $(g,n)$
\begin{equation}
\begin{aligned}
 Z_{g,n}(\bt_1,\cdots,\bt_n)&=\int_C\left(\prod_{i=1}^n\frac{dz_i}{2\pi\ri}
e^{\frac{\bt_i}{2\ga}z_i^2}\right)
W_{g,n}(z_1,\cdots,z_n),
\end{aligned} 
\label{eq:Z-W}
\end{equation}
where $W_{g,n}$ is the genus $g$ part of the connected correlator
of $n$ resolvents.
$W_{g,n}$ satisfies the topological recursion
\begin{equation}
\begin{aligned}
 W_{g,n+1}(z_0,J)=\underset{z=0}{\text{Res}}\,K(z_0,z)
\Bigl[W_{g-1,n+2}(z,-z,J)+\sum_{h=0}^g\sum'_{I\subset J}
W_{h,1+|I|}(z,I)W_{g-h,1+n-|I|}(-z,J\backslash I)\Bigr],
\end{aligned} 
\label{eq:top-rec2}
\end{equation}
with the initial condition
\begin{equation}
\begin{aligned}
W_{0,2}(z_1,z_2)=\frac{1}{(z_1-z_2)^2}.
\end{aligned} 
\end{equation}
The recursion kernel is
\begin{equation}
\begin{aligned}
 K(z_0,z)=\frac{1}{(z_0^2-z^2)4y(z)}.
\end{aligned} 
\end{equation}
For the spectral curve $(x(z),y(z))$ in \eqref{eq:y-sin}
\begin{equation}
\begin{aligned}
 x(z)=z^2,\quad y(z)=\frac{\sin(2\pi z)}{4\pi}, 
\end{aligned} 
\end{equation}
it was shown in \cite{Eynard:2007fi} that
$W_{g,n}$ is related to the Weil-Petersson volume $V_{g,n}$ 
\begin{equation}
\begin{aligned}
 W_{g,n}(z_1,\cdots,z_n)=\int_0^\infty \left(\prod_{i=1}^n b_idb_i e^{-b_iz_i}\right)
V_{g,n}(b_1,\cdots,b_n).
\end{aligned} 
\label{eq:W-V}
\end{equation}
Plugging \eqref{eq:W-V} into \eqref{eq:Z-W},
we find that $Z_{g,n}$ is written as
\begin{equation}
\begin{aligned}
 Z_{g,n}(\bt_1,\cdots,\bt_n)
&=\int_0^\infty \left(\prod_{i=1}^n b_idb_i Z_{\text{trumpet}}(\bt_i,b_i)\right)
V_{g,n}(b_1,\cdots,b_n),
\end{aligned} 
\label{eq:Zgn-JT}
\end{equation}
where $Z_{\text{trumpet}}(\bt,b)$ is the trumpet partition function
of JT gravity
\begin{equation}
\begin{aligned}
 Z_{\text{trumpet}}(\bt,b)=\int_{-\infty}^\infty\frac{dk}{2\pi}
e^{-\ri  bk-\frac{\bt}{2\ga}k^2}=\rt{\frac{\ga}{2\pi\bt}}
e^{-\frac{\ga b^2}{2\bt}}.
\end{aligned} 
\label{eq:trumpet-JT}
\end{equation}

\subsection{Triple scaling limit of DSSYK}
As discussed in \cite{Berkooz:2018jqr,Lin:2022rbf},
DSSYK reduces to JT gravity in the  
the triple scaling limit
\begin{equation}
\begin{aligned}
 \la\to0,~ \th\to0\quad\text{with}\quad k=\frac{\th}{\la}:~\text{fixed}.
\end{aligned} 
\label{eq:tri-lim}
\end{equation}
This corresponds to the double-scaling limit of the ETH matrix model,
zooming in on the low energy end $E=-E_0$ of the eigenvalue spectrum
$E(\th)=-E_0\cos\th$.

Let us first see this limit at the disk level.
Using the expansion of $\mu(\th)$ obtained in \cite{Okuyama:2025fhi}
\begin{equation}
\begin{aligned}
 \mu(\th)
=2 q^{-\frac{1}{8}}\rt{\frac{2\pi}{\la}} 
\sin\th e^{-\frac{2\th^2}{\la}}\sum_{j=0}^\infty (-1)^j
e^{-\frac{2\th_j^2}{\la}}2\sinh\left(\frac{4\th\th_j}{\la}\right),
\end{aligned} 
\end{equation} 
with $\th_j=\pi(j+1/2)$, we find that 
$\mu(\th)$ reduces to the Schwarzian density $\rho_s(k)$ in 
\eqref{eq:rho-s} in the triple scaling limit \eqref{eq:tri-lim}
\begin{equation}
\begin{aligned}
 L\mu(\th)\frac{d\th}{2\pi}\to e^{S_0} \rho_s(k)\frac{dk}{\pi},
\end{aligned} 
\label{eq:lim-rhos}
\end{equation}
where $e^{S_0}$ is given by
\begin{equation}
\begin{aligned}
 e^{S_0}=2L(2\pi\la)^{\frac{3}{2}}e^{-\frac{\pi^2}{2\la}}.
\end{aligned} 
\label{eq:S0-disk}
\end{equation}
From the expansion of $E(\th)$
\begin{equation}
\begin{aligned}
E(\th)=-E_0\cos(\la k)\approx -E_0+\hf E_0\la^2k^2,
\end{aligned} 
\label{eq:exp-Eth}
\end{equation}
the disk partition function \eqref{eq:disk}
of DSSYK reduces to 
that of JT gravity \eqref{eq:Z-JT2}
\begin{equation}
\begin{aligned}
 L\cZ_{0,1}(\bt)&\approx 
e^{S_0}\int_{-\infty}^\infty \frac{dk}{2\pi} 
\rho_s(k)e^{\bt E_0-\hf\bt E_0\la^2 k^2}
=e^{S_0+\bt E_0}Z_{\text{JT}}(\bt),
\end{aligned} 
\label{eq:lim-ZJT}
\end{equation}
where $\ga$ is given by
\begin{equation}
\begin{aligned}
 \ga=\frac{1}{E_0\la^2}.
\end{aligned} 
\label{eq:ga-E0}
\end{equation}
Originally, the range of $k$  in \eqref{eq:lim-rhos}
is $k\in[0,\infty]$,
but in \eqref{eq:lim-ZJT} we extended the range of $k$ to
$[-\infty,\infty]$ and divided by 2.

One can also check that $\cZ_{g,n}$ of DSSYK reduces to
$Z_{g,n}$ of JT gravity in the triple scaling limit \eqref{eq:tri-lim}. 
To see this, let us consider the limit of trumpet
in \eqref{eq:trumpet-DSSYK}.
In addition to \eqref{eq:tri-lim}, we take the following limit
\begin{equation}
\begin{aligned}
 \la\to0,~\fb\to\infty\quad \text{with}\quad b=\la\fb:~\text{fixed}.
\end{aligned} 
\label{eq:b-lim}
\end{equation}
Then the trumpet of DSSYK reduces to that of JT gravity in \eqref{eq:trumpet-JT}
\begin{equation}
\begin{aligned}
I_\fb(\bt E_0)=\int_{-\pi}^\pi\frac{d\th}{2\pi}e^{\ri\fb\th}e^{-\bt E(\th)}
&\approx
\int_{-\infty}^\infty \frac{\la dk}{2\pi}e^{\ri bk}e^{\bt E_0-\hf \bt E_0\la^2k^2}\\
&=\la e^{\bt E_0}Z_{\text{trumpet}}(\bt,b),
\end{aligned} 
\label{eq:lim-trumpet}
\end{equation}
where $\ga$ is given by \eqref{eq:ga-E0}.
From the structure of the topological recursion \eqref{eq:top-rec},
it is natural to decompose
\begin{equation}
\begin{aligned}
 N_{g,n}(\fb_1,\cdots,\fb_n)=N_{g,n}^{+}(\fb_1,\cdots,\fb_n)+
N_{g,n}^{-}(\fb_1,\cdots,\fb_n),
\end{aligned} 
\end{equation}
where $N^\pm_{g,n}$ correspond to the residue at $z=\pm1$, respectively.
For even $\fb_i$, we observe that
$N^{+}_{g,n}$ and $N^{-}_{g,n}$ are equal 
\begin{equation}
\begin{aligned}
 N_{g,n}^{+}(\fb_1,\cdots,\fb_n)=N_{g,n}^{-}(\fb_1,\cdots,\fb_n),\quad
(\fb_i\in 2\mathbb{Z}_{+}).
\end{aligned} 
\end{equation}
It was also observed in \cite{Okuyama:2023kdo}
that $N^{\pm}_{g,n}$ reduces to the Weil-Petersson volume
$V_{g,n}$ in the limit \eqref{eq:tri-lim} and \eqref{eq:b-lim}
\begin{equation}
\begin{aligned}
 N_{g,n}^{\pm}(\fb_1,\cdots,\fb_n)\to 
\cN^{2-2g-n}\la^{n}V_{g,n}(b_1,\cdots,b_n),
\end{aligned}
\label{eq:lim-N} 
\end{equation}
where
\begin{equation}
\begin{aligned}
 \cN=2(q;q)_\infty^3\la^3\approx 2(2\pi\la)^{\frac{3}{2}}e^{-\frac{\pi^2}{2\la}}.
\end{aligned} 
\end{equation}
Using \eqref{eq:lim-trumpet}
and \eqref{eq:lim-N}, we find that 
the contribution of $\cZ_{g,n}$ from $z=1$ reduces to 
$Z_{g,n}$ of JT gravity in \eqref{eq:Zgn-JT}
\begin{equation}
\begin{aligned}
 &L^{2-2g-n}\sum_{\fb_1,\cdots,\fb_n\in\mathbb{Z}_{+}}
\left(\prod_{i=1}^n \fb_i I_{\fb_i}(\bt E_0)\right)N^{+}_{g,n}(\fb_1,\cdots,\fb_n)\\
\to & ~e^{-(2g-2+n)S_0}
\int_0^\infty\left(\prod_{i=1}^n b_idb_iZ_{\text{trumpet}}(\bt_i,b_i)
e^{\bt_i E_0}\right)
V_{g,n}(b_1,\cdots,b_n)\\
=&~e^{-(2g-2+n)S_0+\sum_{i=1}^n\bt_iE_0}Z_{g,n}(\bt_1,\cdots,\bt_n)
\end{aligned} 
\end{equation}
where
\begin{equation}
\begin{aligned}
 e^{S_0}=L\cN.
\end{aligned} 
\end{equation}
Interestingly, this agrees with $e^{S_0}$ determined at the disk level 
\eqref{eq:S0-disk}.

\section{Scaling limit around $E=E_0$}\label{sec:dS-JT}
Next, let us consider the limit where we zoom in on 
the high energy end $E=E_0$ of the spectrum $E(\th)$, 
which corresponds to $\th=\pi$ instead of $\th=0$.
In a similar manner as \eqref{eq:tri-lim},
it is natural to take the triple scaling limit
\begin{equation}
\begin{aligned}
 \la\to0,~\th\to\pi\quad\text{with}\quad k=\frac{\pi-\th}{\la}:
~\text{fixed}.
\end{aligned}
\label{eq:tri2} 
\end{equation}
Due to the symmetry in \eqref{eq:Emu-sym},
$\mu(\th)$ reduces to the Schwarzian density $\rho_s(k)$ in this limit 
\eqref{eq:tri2} as well.
However, $E(\th)$ changes sign under $\th\to\pi-\th$
and hence the expansion of $E(\th)$ for $\th=\pi-\la k$ 
has the opposite sign compared to the expansion around $\th=0$
in \eqref{eq:exp-Eth}
\begin{equation}
\begin{aligned}
 E(\pi-\la k)\approx E_0-\hf E_0\la^2k^2.
\end{aligned} 
\end{equation}
This leads to a wrong sign Gaussian integral for $\bt>0$
\cite{Okuyama:2025fhi}
\begin{equation}
\begin{aligned}
 L\cZ_{0,1}(\bt)\to \int_{-\infty}^\infty\frac{dk}{2\pi}\rho_s(k)
e^{S_0-\bt E_0+\hf \bt E_0\la^2k^2},
\end{aligned} 
\label{eq:wrong}
\end{equation}
which indicates that $\th=\pi$ is an unstable saddle point.
In order to make this integral convergent, we have to rotate the 
integration contour to the imaginary direction
\begin{equation}
\begin{aligned}
 k=-\ri\til{k},\quad (\til{k}\in\mathbb{R}).
\end{aligned} 
\label{eq:rotate}
\end{equation}
Then we find
\begin{equation}
\begin{aligned}
 L\cZ_{0,1}(\bt)&\to \ri\int_{-\infty}^\infty \frac{d\til{k}}{2\pi}
\frac{\til{k}\sin(2\pi\til{k})}{2\pi}
e^{S_0-\bt E_0-\hf \bt E_0\la^2\til{k}^2}\\
&=\ri e^{S_0-\bt E_0}\til{Z}_{0,1}(-\bt)
\end{aligned} 
\end{equation}
where $\til{Z}_{0,1}(\til{\bt})$ is given by
\begin{equation}
\begin{aligned}
 \til{Z}_{0,1}(\til{\bt})=\frac{1}{\rt{2\pi(-\til{\bt}/\ga)^3}}
e^{\frac{2\pi^2\ga}{\til{\bt}}},\quad 
(\til{\bt}=-\bt<0),
\end{aligned} 
\label{eq:tilZ01}
\end{equation}
and $\ga$ is defined in \eqref{eq:ga-E0}.
Interestingly, 
$\til{Z}_{0,1}(\til{\bt})$ in \eqref{eq:tilZ01}
is exactly the disk amplitude of de Sitter JT gravity computed in 
\cite{Cotler:2024xzz}.
This suggests that de Sitter JT gravity appears 
from the expansion of ETH matrix model around the unstable saddle point $E=E_0$,
while the stable saddle point $E=-E_0$ corresponds to
the AdS JT gravity (see \eqref{eq:rho-fig}).
Note that $Z_{0,1}(\bt)$ of AdS JT gravity \eqref{eq:Z-JT} and
$\til{Z}_{0,1}(\til{\bt})$ in \eqref{eq:tilZ01} are related by
\begin{equation}
\begin{aligned}
Z_{0,1}(\til{\bt})=\ri 
\til{Z}_{0,1}(\til{\bt}),\quad (\til{\bt}\in\mathbb{R}_{<0}).
\end{aligned} 
\end{equation}
In order to generalize this relation to higher genus, we note that $\til{Z}_{0,1}(\til{\bt})$
is written as
\begin{equation}
\begin{aligned}
 Z_{0,1}(\til{\bt})=\ri\int_{\til{C}}\frac{d\til{z}}{2\pi\ri}
W_{0,1}(\ri\til{z})e^{-\frac{\til{\bt}}{2\ga}\til{z}^2},
\end{aligned} 
\label{eq:Z-tilbt}
\end{equation}
where the integration contour $\til{C}$ is along the imaginary axis
with the opposite orientation from $C$ in \eqref{eq:Z01}
\begin{equation}
\begin{aligned}
 \til{C}=-\ri\mathbb{R},\quad\int_{\til{C}}d\til{z}=\int_{\ri\infty}^{-\ri\infty}
d\til{z}.
\end{aligned} 
\end{equation}
The relation \eqref{eq:Z-tilbt} can be generalized as
\begin{equation}
\begin{aligned}
 Z_{g,n}(\til{\bt}_1,\cdots,\til{\bt}_n)=
\ri^n\int_{\til{C}}\left(\prod_{i=1}^n\frac{d\til{z}_i}{2\pi\ri} 
e^{-\frac{\til{\bt}_i}{2\ga}\til{z}_i^2}\right)
W_{g,n}(\ri\til{z}_1,\cdots,\ri\til{z}_n).
\end{aligned} 
\label{eq:Zgn-tilbt}
\end{equation}

Following \cite{Cotler:2024xzz}, 
we define $\til{W}_{0,1}(\til{z})$ and $\til{y}(\til{z})$ by
\begin{equation}
\begin{aligned}
\til{W}_{0,1}(\til{z})= W_{0,1}(\ri\til{z})=-2\til{z}\til{y}(\til{z}),\quad
\til{y}(\til{z})=-\frac{\sinh(2\pi\til{z})}{2\pi}.
\end{aligned} 
\label{eq:tily}
\end{equation}
Then \eqref{eq:Z-tilbt} becomes
\begin{equation}
\begin{aligned}
 Z_{0,1}(\til{\bt})=\ri 
\int_{\til{C}}\frac{ d\til{z}}{2\pi\ri}
\til{W}_{0,1}(\til{z})e^{-\frac{\til{\bt}}{2\ga}\til{z}^2}
=\ri\til{Z}_{0,1}(\til{\bt}).
\end{aligned} 
\label{eq:tZ01-tW01}
\end{equation}
We also define $\til{W}_{g,n}$ by the topological recursion with
respect to the spectral curve $\til{y}(\til{z})$ in \eqref{eq:tily}.
As shown in \cite{Cotler:2024xzz},
$W_{g,n}$ and $\til{W}_{g,n}$ are related by
\begin{equation}
\begin{aligned}
 W_{g,n}(\ri\til{z}_1,\cdots\ri\til{z}_n)=(-1)^{3g-3+n}
\til{W}_{g,n}(\til{z}_1,\cdots,\til{z}_n),
\end{aligned} 
\label{eq:W-tW}
\end{equation}
which can be proved inductively using the topological recursion.
Thus we find
\begin{equation}
\begin{aligned}
 Z_{g,n}(\til{\bt}_1,\cdots,\til{\bt}_n)&=\ri^n(-1)^{3g-3+n}
 \int_{\til{C}}\left(\prod_{i=1}^n\frac{d\til{z}_i}{2\pi\ri} 
e^{-\frac{\til{\bt}_i}{2\ga}\til{z}_i^2}\right)
\til{W}_{g,n}(\til{z}_1,\cdots,\til{z}_n)\\
&=e^{\frac{3\pi\ri}{2}(2g-2+n)}\til{Z}_{g,n}(\til{\bt}_1,\cdots\til{\bt}_n),
\end{aligned} 
\label{eq:tZgn-tWgn}
\end{equation}
where $\til{Z}_{g,n}$ is defined in a similar manner as \eqref{eq:Z-W}
\begin{equation}
\begin{aligned}
 \til{Z}_{g,n}(\til{\bt}_1,\cdots\til{\bt}_n)=\int_{\til{C}}
\left(\prod_{i=1}^n\frac{d\til{z}_i}{2\pi\ri} 
e^{-\frac{\til{\bt}_i}{2\ga}\til{z}_i^2}\right)
\til{W}_{g,n}(\til{z}_1,\cdots,\til{z}_n).
\end{aligned} 
\label{eq:tZ-tW}
\end{equation}
One can see that \eqref{eq:tZ01-tW01} for $(g,n)=(0,1)$ 
is a special case of \eqref{eq:tZgn-tWgn}
for general $(g,n)$.
Finally, the genus expansion becomes
\begin{equation}
\begin{aligned}
 \sum_{g=0}^\infty e^{-(2g-2+n)S_0}
Z_{g,n}(\til{\bt}_1,\cdots,\til{\bt}_n)
=\sum_{g=0}^\infty e^{-(2g-2+n)\til{S}_0}\til{Z}_{g,n}(\til{\bt}_1,\cdots\til{\bt}_n)
\end{aligned} 
\end{equation}
with
\begin{equation}
\begin{aligned}
 e^{\til{S}_0}=e^{S_0-\frac{3\pi\ri}{2}}=\ri e^{S_0}.
\end{aligned} 
\label{eq:def-tilS0}
\end{equation}
This reproduces the relation between $\til{S}_0$ and $S_0$
found in \cite{Cotler:2024xzz}.

Essentially, $\til{Z}_{g,n}$ is obtained from $Z_{g,n}$
by the analytic continuation $\bt\to-\bt$.
In fact, the one-point function of $\Tr e^{-\bt H}$ in the 
ETH matrix model
is an even function of $\bt$ and hence it is expanded as
\begin{equation}
\begin{aligned}
 \bra\Tr e^{-\bt H}\ket\approx
\sum_{g=0}^\infty e^{-(2g-1)S_0}
\bigl[e^{\bt E_0}Z_{g,1}(\bt)+e^{-\bt E_0}Z_{g,1}(-\bt)\bigr],
\end{aligned} 
\label{eq:one-pt}
\end{equation}
where the first term comes from $\th=0$ while the second term comes from
$\th=\pi$ \cite{Okuyama:2025fhi}.
Here we have ignored the contribution of $\th\ne0,\pi$ for simplicity.
For $\bt>0$, the second term of \eqref{eq:one-pt}
contains the imaginary part due to the expansion around the 
unstable saddle point $\th=\pi$.
After absorbing the imaginary part into $\til{S}_0$ in \eqref{eq:def-tilS0},
the one-point function \eqref{eq:one-pt} becomes
\begin{equation}
\begin{aligned}
 \bra\Tr e^{-\bt H}\ket\approx
\sum_{g=0}^\infty
\bigl[e^{-(2g-1)S_0+\bt E_0}Z_{g,1}(\bt)
+e^{-(2g-1)\til{S}_0+\til{\bt} E_0}\til{Z}_{g,1}(\til{\bt})\bigr],
\end{aligned} 
\label{eq:Tr-sum}
\end{equation}
with $\til{\bt}=-\bt$. 
In the computation of the second term, we have to rotate the contour 
as in \eqref{eq:rotate}, which can be nicely summarized by the replacement 
$W_{g,n}\to\til{W}_{g,n}$
in \eqref{eq:Z-W} and \eqref{eq:tZ-tW}.
From the limit $N^\pm_{g,n}\to V_{g,n}$ in \eqref{eq:lim-N},
it is guaranteed that $\om_{g,n}$ reduces to
$W_{g,n}$ and $\til{W}_{g,n}$ in the scaling limit around
$z=1$ and $z=-1$.
Since $W_{g,n}$ and $\til{W}_{g,n}$ are related by \eqref{eq:W-tW},
the only difference between $Z_{g,n}$ and $\til{Z}_{g,n}$ is 
the integration contour.

Note that the eigenvalue density $\rho_0(E)$ in \eqref{eq:rho-fig} is symmetric
under $E\to -E$ since the potential $V(H)$ of the ETH matrix model in \eqref{eq:VH}
is an even function of $H$.
However, the Boltzmann factor $e^{-\bt E}$ breaks this symmetry $E\to-E$. 
As a consequence, the two saddle points $E=-E_0$ and $E=E_0$ become distinct:
the former corresponds to $AdS_2$ while the latter corresponds to $dS_2$ 
as depicted in \eqref{eq:rho-fig}.

\subsection{Hartle-Hawking wavefunction}
As discussed in \cite{Maldacena:2019cbz}, $\til{Z}_{0,1}(\til{\bt})$
for imaginary $\til{\bt}$ is interpreted as the Hartle-Hawking wavefunction
of $dS_2$.
Following \cite{Cotler:2024xzz}, we set
\begin{equation}
\begin{aligned}
 \til{\bt}=-\ri\ell-\vep,
\end{aligned} 
\end{equation}
where we introduced a small positive $\vep$ to make the real part of $\til{\bt}$
slightly negative.
It is also convenient to set
\begin{equation}
\begin{aligned}
 E_0=\frac{1}{\la},\quad \ga=\frac{1}{E_0\la^2}=\frac{1}{\la}.
\end{aligned} 
\label{eq:E0-la}
\end{equation}
Then the disk amplitude $\cZ_{0,1}(\bt)$ of the ETH matrix model near $\th=\pi$
in the scaling limit \eqref{eq:tri2} becomes
\begin{equation}
\begin{aligned}
 L\cZ_{0,1}(\bt)&\approx e^{\til{S}_0+\til{\bt}E_0}\til{Z}_{0,1}(\til{\bt})\\
&=\frac{1}{\rt{2\pi(-\til{\bt}\la)^3}}e^{\til{S}_0+
\frac{\til{\bt}}{\la}+\frac{2\pi^2}{\til{\bt}\la}}\\
&=\frac{1}{\rt{2\pi(\ri\ell\la)^3}}e^{\til{S}_0+\frac{1}{\la}\bigl(-\ri\ell+\ri\frac{2\pi^2}{\ell}\bigr)}.
\end{aligned} 
\label{eq:HH}
\end{equation}
This agrees with the Hartle-Hawking wavefunction $\Psi_+$
in \cite{Maldacena:2019cbz} if we identify the boundary value $\phi_b$
of the dilaton as (see (2.12) in \cite{Maldacena:2019cbz})
\begin{equation}
\begin{aligned}
 2\phi_b=\frac{1}{\la}.
\end{aligned} 
\end{equation}
It is interesting that the linear term of $\ell$ in the exponent
of $\Psi_+$ in \cite{Maldacena:2019cbz} 
is correctly reproduced from the Boltzmann factor
$e^{\til{\bt}E_0}$.

\section{Classical solutions of the sine dilaton gravity}\label{sec:sine-dilaton}
In this section we will argue that the de Sitter interpretation of 
the ETH matrix model near $E=E_0$ is consistent with the sine dilaton gravity,
at least at the classical level.
However, we emphasize that 
the de
Sitter interpretation of the upper edge $E=E_0$ 
holds fully quantum mechanically 
and not just classically in the context of sine dilaton gravity.

The action of the sine dilaton gravity is given by
\begin{equation}
\begin{aligned}
 S=-\frac{1}{2\la}\int d^2x\rt{g}\bigl[\Phi R+W(\Phi)\bigr],\quad
W(\Phi)=2\sin\Phi.
\end{aligned} 
\label{eq:sine-dilaton}
\end{equation}
For small $\Phi$, this action reduces to the JT gravity with 
a negative cosmological constant
\begin{equation}
\begin{aligned}
 S=-\frac{1}{2\la}\int d^2x\rt{g}\Phi(R+2).
\end{aligned} 
\end{equation}
On the other hand, for $\Phi=\pi-\til{\Phi}$ with small 
$\til{\Phi}$, the action \eqref{eq:sine-dilaton} becomes
\begin{equation}
\begin{aligned}
 S=\frac{1}{2\la}\int d^2x\rt{g}\til{\Phi}(R-2),
\end{aligned} 
\end{equation}
up to an additive constant. This is the action of 
JT gravity with a positive cosmological constant.
As discussed in \cite{Blommaert:2024whf},
this observation suggests that the sine dilaton gravity
contains some information of de Sitter JT gravity.
We will check this expectation at the level of classical solutions.

To find a classical solution of the
dilaton gravity with the general potential
$W(\Phi)$, it is convenient to 
choose the gauge
where $\Phi$ is identified with the radial coordinate $r$
\cite{Witten:2020ert,Gegenberg:1994pv}
\begin{equation}
\begin{aligned}
 ds^2=A(r)d\tau^2+\frac{dr^2}{A(r)},\quad \Phi(r)=r.
\end{aligned} 
\label{eq:sol}
\end{equation}
Then $A(r)$ is determined by the condition
\begin{equation}
\begin{aligned}
 A'(r)=W(r).
\end{aligned} 
\end{equation}
For the sine dilaton case $W(r)=2\sin r$, the classical solution for $A(r)$
is given by \cite{Blommaert:2024ymv}
\begin{equation}
\begin{aligned}
 A(r)=2\cos\th-2\cos r,
\end{aligned} 
\label{eq:Ar}
\end{equation}
where $\th$ is an integration constant which parameterizes the classical solution. 
The ADM energy of this solution is \cite{Mertens:2022irh,Blommaert:2024ymv}
\begin{equation}
\begin{aligned}
 E=-\frac{1}{\la}\cos\th,
\end{aligned} 
\end{equation}
which agrees with the energy spectrum $E(\th)$ of DSSYK
in \eqref{eq:Eth-muth} under the identification $E_0=\la^{-1}$ in
\eqref{eq:E0-la}. 

Let us expand $A(r)$ around $r=\th$. Setting $u=r-\th$ we find
\begin{equation}
\begin{aligned}
 A(r)=2 u \sin\th+u^2\cos\th +\cO(u^3). 
\end{aligned} 
\label{eq:A-expand}
\end{equation}
The $\cO(u^1)$ term is non-zero for $\th\ne0,\pi$;
in this case the classical solution \eqref{eq:sol}
describes a Euclidean black hole with the horizon at $r=\th$.
However, when $\th=0,\pi$, the $\cO(u^1)$ term
vanishes and $A(r)$ starts at $\cO(u^2)$
\begin{equation}
\begin{aligned}
 A(r)=\pm u^2,
\end{aligned} 
\end{equation}
where the plus and minus signs correspond to $\th=0$ and $\th=\pi$, respectively. Then the metric in \eqref{eq:sol} becomes
\begin{equation}
\begin{aligned}
 ds^2=\pm\left(u^2d\tau^2+\frac{du^2}{u^2}\right).
\end{aligned} 
\label{eq:met-0pi}
\end{equation}
The metric inside the parenthesis 
is nothing but the metric of Euclidean $AdS_2$ in the Poincar\'{e} 
coordinate.
This is consistent with the fact that
JT gravity with a negative cosmological constant is reproduced
from the triple scaling limit of DSSYK near $\th=0$.
For $\th=\pi$, the metric in \eqref{eq:met-0pi} is the minus
of the metric of $AdS_2$, which is called $-AdS_2$ in \cite{Maldacena:2019cbz}.
As explained in \cite{Maldacena:2019cbz}, the metric of $-AdS_2$
has a constant positive curvature and hence it can be thought of as a
two-dimensional de Sitter space
\begin{equation}
\begin{aligned}
 dS_2=-AdS_2.
\end{aligned} 
\end{equation}
This is consistent with our identification of the 
triple scaling limit of DSSYK near $\th=\pi$ as the de Sitter JT gravity.

Note that the disk partition function
of DSSYK is approximated by a sum over the
classical solutions of the sine dilaton gravity
\eqref{eq:sol} and \eqref{eq:Ar}
parameterized by $\th$
\begin{equation}
\begin{aligned}
 \cZ_{0,1}(\bt)=\int_0^\pi\frac{d\th}{2\pi}\mu(\th)
e^{-\bt E(\th)}\approx\sum_{\text{classical~sol.}}e^{-\bt E(\th)}.
\end{aligned} 
\label{eq:sum-sol}
\end{equation}
This suggests that the contributions of $\th=0$ and $\th=\pi$ are
different saddle points of the bulk action.
We should stress that this picture is different from the 
holographic description of the de Sitter bubble inside AdS
using the AdS/CFT correspondence \cite{Freivogel:2005qh,Freivogel:2006xu}.
In our case,
$dS_2$ at $\th=\pi$
is not a bubble inside $AdS_2$ at $\th=0$; they are 
distinct solutions of the bulk action
and hence they are not interpreted as two regions of a single spacetime
separated by some kind of wall.
Rather, they are simply added 
as two independent contributions to $\bra\Tr e^{-\bt H}\ket$,
as shown in \eqref{eq:Tr-sum}.

\section{Conclusion and outlook}\label{sec:conclusion}
In this paper, we found that the triple scaling limit 
\eqref{eq:tri2} of the ETH matrix model
around the upper end $E=E_0$ reproduces the de Sitter
JT gravity studied in \cite{Cotler:2024xzz}.
Our identification of de Sitter JT gravity is also supported by the
behavior \eqref{eq:met-0pi}
of the classical solution of the sine dilaton gravity.
In our construction, the de Sitter JT gravity is embedded in a more complete theory, 
i.e., the ETH matrix model, and appears as an unstable saddle point 
of the ETH matrix model.

There are many interesting open questions.
In this work, we have ignored the effect of matter operators for
simplicity.
As discussed in \cite{Jafferis:2022wez}, if we include the 
effect of matter operator, the ETH matrix model becomes a two-matrix model.
It would be interesting to study the behavior of this two-matrix model around
the unstable saddle point $E=E_0$ along the lines of \cite{Okuyama:2023aup,Okuyama:2023yat}. 

We did not introduce the observer in our construction
of de Sitter JT gravity. It is argued that the observer is important to define
the quantum gravity on de Sitter space 
\cite{Chandrasekaran:2022cip,Witten:2023qsv}.
It would be interesting to understand the role of observer in our construction
of de Sitter JT gravity. Perhaps, the matter operators
play the role of observers.

The eigenvalue density $\rho_0(E)$ of the ETH matrix model 
has a square-root branch cut near the edge $E=\pm E_0$ of the spectrum
\begin{equation}
\begin{aligned}
 \rho_0(E)\sim\rt{E_0^2-E^2}.
\end{aligned} 
\end{equation}
However, this square-root branch cut exists only in the strict large $L$ limit.
At finite $L$ the square-root branch cut disappears and the 
exact finite $L$ result drastically differs from the large $L$
limit, as discussed in \cite{Maldacena:2004sn} in
the case of minimal string theory.
It would be interesting to understand the fate of our $dS_2$ saddle 
at finite $L$.
It is argued in \cite{Witten:2001kn}
that the dimension of the Hilbert space of de Sitter quantum gravity
is finite, determined by the Gibbons-Hawking entropy 
\cite{Gibbons:1977mu}. It would be interesting to understand the 
Hilbert space of the ETH matrix model at finite $L$ \cite{Miyaji:2025ucp}.

Our identification of $dS_2$ in DSSYK is different from the
proposal in Susskind \textit{et al.} \cite{Susskind:2021esx,Susskind:2022dfz,Susskind:2022bia,Susskind:2023hnj,Rahman:2023pgt,Rahman:2024iiu,Sekino:2025bsc} 
or Verlinde \textit{et al.} \cite{Narovlansky:2023lfz,Verlinde:2024znh,Verlinde:2024zrh,Tietto:2025oxn}:
they seem to identify the de Sitter space as a contribution
of $E=0$, i.e. the middle part of the support of $\rho_0(E)$ in 
\eqref{eq:rho-fig}, not the upper end $E=E_0$.  
It would be interesting to understand the relationship between their
proposal and ours, if any.

\acknowledgments
The author would like to thank Tokiro Numasawa for correspondence.
This work was supported
in part by JSPS Grant-in-Aid for Transformative Research Areas (A) 
``Extreme Universe'' 21H05187 and JSPS KAKENHI 25K07300.

\bibliography{paper}
\bibliographystyle{utphys}

\end{document}